\documentclass[fleqn,twoside]{article}
\usepackage{espcrc2}

\usepackage{graphicx}
\usepackage[figuresright]{rotating}

\newcommand{\AmS}{{\protect\the\textfont2
  A\kern-.1667em\lower.5ex\hbox{M}\kern-.125emS}}

\title{Hybrid mesons from anisotropic lattice QCD with the clover and improved gauge actions}

\author{Xiang-Qian Luo\address[ZSU]{Department of Physics, Zhongshan (Sun Yat-Sen) University, 
Guangzhou 510275, China}%
        \thanks{Email: stslxq@zsu.edu.cn. Work supported by 
National Science Fund for Distinguished Young Scholars, National Science Foundation,
Guangdong Provincial Natural Science Foundation, 
Ministry of Education, and Foundation of Zhongshan Univ. Advanced Research Center.}
      and  Zhong-Hao Mei\addressmark
}

\begin{document}

\begin{abstract}
We study hybrid mesons from the clover and improved gauge actions at $\beta=2.6$ on the 
anisotropic $12^3\times36$ lattice  using our PC cluster. We estimate the mass of $1^{-+}$ 
light quark hybrid as well as the mass of the charmonium hybrid. 
The improvement of both quark and gluonic actions, first applied to the hybrid mesons, is shown to be more
efficient in reducing the lattice spacing and finite volume errors. 
\vspace{1pc}
\end{abstract}

\maketitle

\section{INTRODUCTION}

Lattice QCD is the ideal approach 
not only for computing $\bar{q}q$ meson spectrum, 
but also for hybrids and glueballs. However, the lattice technique 
is not free of systematic errors.
The Wilson gauge and quark actions 
suffer from significant lattice spacing errors, 
which are smaller only at very large $\beta$,
and very large lattice volume is required to get rid of finite size effects.

There have been several quenched lattice calculations of hybrid meson masses,  
part of them are listed in Tab.~\ref{tab6}.
In Ref.~\cite{Bernard:1997ib}, 
the Wilson gluon action and quark action were used.
In Refs.~\cite{Lacock:1996ny,McNeile:1998cp}, 
the authors used Wilson gauge action and SW improved quark action. 
For the hybrid mesons containing heavy quarks ${\bar Q}Qg$,
the NRQCD action\cite{Manke:1998qc} and the LBO action\cite{Juge:1999ie} 
have also been applied. 
There is also a recent work using the improved KS quark action\cite{Gregory}.

In this work, we employ {\it both improved gluon and quark actions on
the anisotropic lattice}, which should have smaller systematic errors, 
and should be more efficient in reducing 
the lattice spacing and finite volume effects. We will present
data for the $1^{-+}$ hybrid mass
and the splitting between the $1^{-+}$ hybrid mass and the 
spin averaged S-wave mass for charmonium. Details can be found 
in Ref.~\cite{Mei:2002ip}.

\section{ACTIONS}
\label{secII}

The total lattice action is $S=S_g+S_q$. 
The improved gluonic action $S_g$ is \cite{Morningstar:1997ff,Luo:1998dx}:
\begin{eqnarray}
    S_g &=& -\beta {1 \over \xi} \sum_{x,j<k} \left( {5 \over 3} {P_{j,k} \over u_s^4}
-{1 \over 12} {R_{j,k} \over u_s^6} - {1 \over 12} {R_{k,j} \over u_s^6} \right)
\nonumber \\
&-& \beta \xi \sum_{x,j} \left( {4 \over 3} {P_{j,4} \over u_s^2}
- {1 \over 12} {R_{j,4} \over u_s^4} \right),
\label{S_g}
\end{eqnarray}
where $P$ stands for a $1\times 1$ plaquette and $R$ for  a $2\times 1$ rectangle.
The SW improved action for quarks\cite{Klassen:1998fh,Okamoto:2001jb} is 
\begin{eqnarray}
    S_q &=& \sum_{x}  {\bar \psi} (x)\psi(x)
\nonumber \\
&-& \kappa_t \sum_{x}
[\bar{\psi}(x) (1-\gamma_0)U_{4}(x)
\psi(x+\hat{4}) 
\nonumber \\
&+& {\bar \psi}(x)(1+\gamma_0)
U^{\dag}_{4}(x) \psi (x-\hat{4})]
\nonumber \\
&-& \kappa_s \sum_{x,j} [{\bar \psi}(x) (1-\gamma_j)U_{j}(x) \psi (x+\hat{j})
\nonumber \\
&+& {\bar \psi} (x) (1+\gamma_j)U^{\dag}_{j}(x-\hat{j})\psi(x-\hat{j})]
\nonumber \\
&+& i {\kappa}_s {C}_s^{TI}  \sum_{x,j<k} {\bar \psi} (x)
\sigma_{jk}{\hat F_{jk}}(x)\psi (x) 
\nonumber \\
&+& i \kappa_s C_t^{TI} \sum_{x,j}{\bar \psi} (x)
\sigma_{j4} {\hat F_{j4}} (x) \psi (x),
\label{S_q}
\end{eqnarray}
where ${\hat F}$ stands for the clover-leaf construction\cite{Luo:1996tx}
for the gauge field tensor.
Tadpole improvement is carried out so that the actions are more continuum-like.

\section{SIMULATIONS}
\label{secIV}

On our PC cluster\cite{Cluster,Luo:2002kr,Mei:2002fu},
the SU(3) pure gauge configurations were
generated with the gluon action in Eq. (\ref{S_g}) 
using Cabibbo-Marinari pseudo-heatbath algorithm. The configurations are decorrelated by 
SU(2) sub-group over-relaxations. We calculated the
tadpole parameter $u_s$ self-consistently. 
90 independent gauge configurations at $\beta=2.6$ and $\xi=3$ on the $12^3 \times 36$ lattice
were stored. 
Although such an ensemble is not very big, it is bigger than earlier 
simulations by UKQCD 
and MILC collaborations\cite{Bernard:1997ib,Lacock:1996ny,McNeile:1998cp} on isotropic lattices.

The quark propagator was obtained by inverting the matrix $\Delta$ 
in $S_q=\sum_{x,y} {\bar \psi}(x) \Delta_{x,y} \psi (y)$ in Eq. (\ref{S_q})  
by means of BICGStab algorithm. The residue is of $O(10^{-7})$.
We computed the correlation functions with
various sources and sinks\cite{Mei:2002ip}, at four values of the Wilson hopping parameter
($\kappa_t=0.4119$, 0.4199, 0.4279, 0.4359).

\begin{figure}[htb]
\rotatebox{270}{\includegraphics[width=55mm]{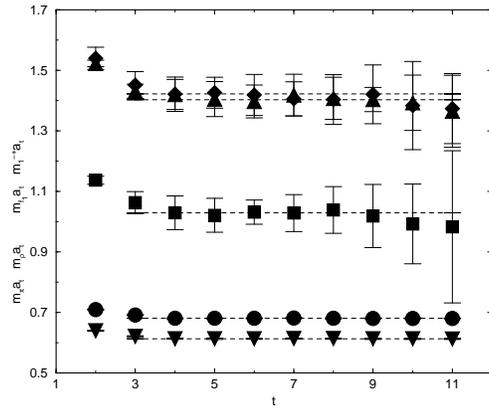}}
\vspace{-1cm}
\caption{\label{mprf1t.eps}  
Effective masses for the $\pi$ (triangle down), $\rho$ (circles), $f_1$ P-wave (square) mesons 
and $1^{-+}$  exotic meson (the diamond for $1^{-+}$ source and the triangle
up for the $q^4$ source).
}
\end{figure}

In Fig.~\ref{mprf1t.eps}, we plot the effective masses for the $\pi$, $\rho$, 
$f_1$ ordinary mesons and 
$1^{-+}$ exotic meson at $\kappa_t=0.4359$. 
For the ordinary mesons, we used the their corresponding operator
as both source and sink. For the exotic meson, we tried two different cases:
(1) the $1^{-+}$ operator as both source and sink; 
(2) the $q^4$ source and $1^{-+}$ sink, which give consistent results within error bars.

The CP-PACS, MILC and UKQCD collaborations 
used the unimproved Wilson gauge action to generate configurations.
They had to work on very large $\beta (> 6$), corresponding to very small $a_s (<$ 0.1~fm), to get rid of the
finite spacing errors.
They had also to use very large lattices $L^3 \ge 20^3$, to avoid strong lattice size effects at such small $a_s$.
In comparison, our lattices are much coarser ($a_s=0.33$~fm), and the number of lattice sites is much smaller.
Our finite size effects could be ignored, for the physical size of the spatial lattice is 
$12^3 a_s^3=(3.96~{\rm fm})^3$
and should be big enough.  
Our results for the effective mass indicate the existence of a much wider 
plateau than in the previous work on isotropic lattices.

\section{RESULTS}
\label{secV} 

By extrapolating the effective mass  of the $1^{-+}$ hybrid meson to the chiral limit, 
and using $a_t$ determined from the $\rho$ mass, we get  $2013 \pm 26 \pm 71$ MeV. 
In Tab.~\ref{tab6}, we compare the results from various lattice methods.
Our result is consistent with the MILC data\cite{Bernard:1997ib}, 
obtained using the Wilson gluon action and clover quark action
on  much larger isotropic lattices and much smaller $a_s$. 

We also show our results in Tab.~\ref{tab6} for the $1^{-+}$ hybrid meson mass in the charm quark sector,
using the method discussed in Refs.~\cite{Bernard:1997ib,McNeile:1998cp}.
Our corresponding $\kappa_t^{\rm charm}=0.1806(5)(18)$ 
is obtained by tuning $(m_{\pi}(\kappa_t \to \kappa_t^{\rm charm})+3m_{\rho} (\kappa_t \to \kappa_t^{\rm charm}))/4
=(m_{\eta_c}+3m_{J/\psi})/4$=3067.6 MeV,
where on the right hand side, the experimental inputs $m_{\eta_c}= 2979.8$ MeV and $m_{J/\psi}=3096.9$ MeV are used.
The $1^{-+}$ hybrid meson mass at our
$1/\kappa_t^{\rm charm}$ is $m_{1^{-+}}= 4369 \pm 37 \pm 99$ MeV, 
is consistent with the MILC data\cite{Bernard:1997ib}.
The splitting between the hybrid meson mass and the spin 
averaged S-wave mass [$m_{1^{-+}}-(m_{\eta_c}+3m_{J/\psi})/4$], at our $\kappa_t^{\rm charm}$ is
 $1302 \pm 37 \pm 99$ MeV, consistent with the CP-PACS data, obtained using the Wilson gluon action and NRQCD quark action
on much larger anisotropic lattices and much smaller $a_s$.

As a byproduct, we give the $f_1$ P-wave $1^{++}$ meson in the chiral limit,
as well as their experimental values\cite{Groom:in}.
If we assume that  the pion is massive and $f_1(1420)$ is made of ${\bar s}s$, 
the  $f_1$ P-wave $1^{++}$ meson mass would be $1499 \pm 28 \pm 65$ MeV.

\section{SUMMARY}

To summarize, we have used the tadpole-improved gluon action and clover action to compute the
hybrid meson masses on much coarser anisotropic lattices. 
The main results are given in Tab.~\ref{tab6} and
compared with other lattice approaches. 
In our opinion, 
our approach is more efficient in reducing systematic errors due to finite lattice spacing.

We would like to thank some CP-PACS, MILC and UKQCD members for useful discussions.

\begin{table*}[htb]
  \begin{tabular}{ccc}\hline
 Light  $1^{-+} {\bar q}qg$  (GeV)     & Method &   Ref.  \\ \hline
   1.97(9)(30) & Isotropic $S_g({\rm W})+S_q({\rm W})$ & MILC97\cite{Bernard:1997ib}   \\
     1.87(20) & Isotropic $S_g^{\rm TI}({\rm W})+S_q^{\rm TI}({\rm SW})$  &UKQCD97\cite{Lacock:1996ny}    \\
   2.11(10) & Isotropic $S_g^{\rm TI}({\rm W})+S_q^{\rm TI}({\rm SW})$ & MILC99\cite{McNeile:1998cp} \\
  2.013(26)(71) & Anisotropic $S_g^{\rm TI}(1\times 1+ 2\times 1)+S_q^{\rm TI}({\rm SW})$  & {\bf ZSU (this work)} \\ \hline \hline
 $1^{-+} {\bar q}qg$  (GeV)      & Method &  Ref. \\ \hline
4.390 (80) (200) & Isotropic $S_g({\rm W})+S_q({\rm W})$ & MILC97\cite{Bernard:1997ib} \\
4.369 (37) (99)  & Anisotropic $S_g^{\rm TI}(1\times 1+ 2\times 1)+S_q^{\rm TI}({\rm SW})$ & {\bf ZSU (this work)} \\ \hline \hline
 $1^{-+} {\bar c}cg$ -1S ${\bar c}c$  splitting (GeV)     & Method &   Ref.  \\ \hline
   1.34(8)(20)& Isotropic $S_g({\rm W})+S_q({\rm W})$ & MILC97\cite{Bernard:1997ib}   \\
    1.22(15) & Isotropic $S_g^{\rm TI}({\rm W})+S_q^{\rm TI}({\rm SW})$ & MILC99\cite{McNeile:1998cp}   \\
    1.323(13) & Anisotropic $S_g^{\rm TI}({\rm W})+S_q^{\rm TI}({\rm NRQCD})$ & CP-PACS99\cite{Manke:1998qc}\\
  1.19 & Isotropic $S_g^{\rm TI}(1\times 1+ 2\times 1)+S_q^{\rm TI}({\rm LBO})$ & JKM99\cite{Juge:1999ie}\\
   1.302(37)(99) & Anisotropic $S_g^{\rm TI}(1\times 1+ 2\times 1)+S_q^{\rm TI}({\rm SW})$ & {\bf ZSU (this work)} \\ \hline
                \end{tabular}
\caption{\label{tab6} Predictions for the masses of hybrid mesons. 
Abbreviations: W for Wilson, $1\times 1+ 2\times 1$ for
the plaquette terms plus the rectangle terms,
SW for  Sheikholeslami-Wohlert (Clover), 
TI for tadpole-improved, NRQCD for non-relativistic QCD, and  LBO for leading Born-Oppenheimer.}
\end{table*}

\begin{thebibliography}{9}

\bibitem{Bernard:1997ib}
C.~Bernard {\it et al.},
Phys.\ Rev.\ {\bf D56} (1997) 7039.



\bibitem{Lacock:1996ny}
P.~Lacock, C.~Michael, P.~Boyle and P.~Rowland,
Phys.\ Lett.\ {\bf B401} (1997) 308.


\bibitem{McNeile:1998cp}
C. Bernard {\it et al.},
Nucl.\ Phys.\  {\bf B(Proc. Suppl.)73} (1999) 264.



\bibitem{Manke:1998qc}
T. Manke {\it et al.},
Phys. Rev. Lett. {\bf 82}(1999) 4396.



\bibitem{Juge:1999ie}
K.~Juge, J.~Kuti and C.~Morningstar,
Phys.\ Rev.\ Lett.\  {\bf 82} (1999) 4400.


\bibitem{Gregory}
C. Bernard, {\it et al.}, (MILC Collaboration), these proceedings.



\bibitem{Mei:2002ip}
Z.~H.~Mei and X.~Q.~Luo,
hep-lat/0206012.


\bibitem{Morningstar:1997ff}
C.~Morningstar and M.~Peardon,
Phys.\ Rev.\ {\bf D56} (1997) 4043;



\bibitem{Luo:1998dx}
X.~Q.~Luo, S.~Guo, H.~Kroger and D.~Schutte,
Phys.\ Rev.\ {\bf D59} (1999) 034503.

\bibitem{Klassen:1998fh}
T.~Klassen,
Nucl.\ Phys.\ {\bf B(Proc.\ Suppl.)73} (1999) 918.


\bibitem{Okamoto:2001jb}
M.~Okamoto {\it et al.},
Phys.\ Rev.\ {\bf D65} (2002)  094508.






\bibitem{Luo:1996tx}
X.~Q.~Luo,
Comput.\ Phys.\ Commun.\  {\bf 94} (1996) 119.


\bibitem{Cluster}
X.Q.~Luo,
E. Gregory, J. Yang, Y. Wang, D. Chang and Y. Lin,
hep-lat/0011090;
hep-lat/0107017.



\bibitem{Luo:2002kr}
X.Q.~Luo,
E. Gregory, H. Xi, J. Yang, Y. Wang, D. Chang and Y. Lin,
Nucl. Phys.  {\bf B(Proc. Suppl.)106} (2002) 1046;



\bibitem{Mei:2002fu}
Z.~H.~Mei, X.~Q.~Luo and E.~B.~Gregory,
Chin.\ Phys.\ Lett.\  {\bf 19} (2002)  636.



\bibitem{Groom:in}
D.~Groom {\it et al.},
Eur.\ Phys.\ J.\ {\bf C15} (2000) 1.




\end{thebibliography}
\end{document}